\documentclass[12pt]{article}
\usepackage[dvips]{color}
\usepackage{epsfig}
\usepackage{amsmath}
\usepackage{graphicx}

\begin{document}
\title{\bf Rotating charged hairy black hole in (2+1) dimensions and particle acceleration}
\author{{J. Sadeghi\thanks{Email:
pouriya@ipm.ir},\hspace{1mm}B. Pourhassan \thanks{Email:
b.pourhassan@umz.ac.ir},\hspace{1mm} and H. Farahani\thanks{Email:
h.farahani@umz.ac.ir}}\\
{\small {\em  Young researchers club, Ayatollah Amoli branch, Islamic azad university, Amol, Iran}}} \maketitle
\begin{abstract}
In this paper we construct rotating charged hairy black hole in
(2+1) dimensions for infinitesimal black hole charge and rotation
parameters. Then we consider this black hole as particle accelerator
and calculate the center-of-mass energy of two colliding test
particles near the rotating charged hairy black hole in (2+1)
dimensions. As we expected, the center-of-mass energy has infinite
value.
\end{abstract}

\section{Introduction}
Recently, charged black hole with a scalar hair in (2+1) dimensions
studied in the Ref. [1] and rotating hairy black hole in (2+1)
dimensions studied in the Ref. [2]. The (2+1)-dimensional theories
are toy models to investigate some fundamental ideas to understand
higher dimensional theories because they are easy to study [3].
Also, it is useful to study gauge/gravity dualities [4-7]. These
kinds of black hole recover well known BTZ black hole in (2+1)
dimensions
[8-12].\\
In this work we would like collect Refs. [1, 2] to construct
rotating charged hairy black hole in (2+1) dimensions. In that case
we assume that the electric charge and rotational parameter are
infinitesimal.\\
Then, the main goal of this paper is particle acceleration
mechanism. It has been shown that free particles falling from rest
at infinity outside a rotating black holes may collide with
arbitrarily high center-of-mass (CM) energy and hence rotating black
holes may be considered as a particle accelerator [13, 14]. It is
found that the CM energy of elastic and inelastic scattering of
particles in the gravitational field of static and rotating Kerr
black holes is limited for the static and is unlimited for the
rotating black holes [15]. Several studies indicate that having
infinite CM energy of colliding particles is a generic property of a
rotating black holes [16-27]. Now, we verify this general property
for the rotating charged hairy black hole in (2+1) dimensions.\\
In the next section we construct a black hole in (2+1) dimensions
include electric charge, scalar charge and rotational parameter, and
then in section 3 we obtain field equations and discuss geometric
properties of this black hole. In section 4 we discuss about horizon
structure of rotating charged hairy black hole and obtain event
horizon for small radius limit. In section 5 we study particle
acceleration and obtain CM energy of two colliding particles near
the black hole. In section 6 we discuss about the effective
potential, and finally in section 7 we summarized our results and
give conclusion.
\section{Rotating charged hairy black hole in (2+1) dimensions}
We consider the (2+1)-dimensional gravity with a non-minimally
coupled scalar field and self coupling potential $V(\phi)$, which is
described by the following action,
\begin{equation}\label{s1}
S=\frac{1}{2}\int{d^{3}
x\sqrt{-g}[R-g^{\mu\nu}\nabla_{\mu}\phi\nabla_{\nu}\phi-\xi
R\phi^2-2V(\phi)-\frac{1}{4}F_{\mu\nu}F^{\mu\nu}]},
\end{equation}
where $\xi$ is a coupling constant between gravity and the scalar
field which will be fixed as $\xi=1/8$. The proposed metric for this
black hole is similar to the Ref. [2],
\begin{eqnarray}\label{s2}
ds^{2}=-f(r)dt^{2}+\frac{1}{f(r)}dr^{2}+r^{2}(d\psi+\omega(r)dt)^{2}.
\end{eqnarray}
Under assumption of infinitesimal $a$ and $Q$ we obtain,
\begin{equation}\label{s3}
f(r)=3\beta-\frac{Q^2}{4}+(2\beta-\frac{Q^2}{9})\frac{B}{r}-Q^2(\frac{1}{2}+\frac{B}{3r})\ln(r)+\frac{(3r+2B)^{2}a^{2}}{r^{4}}+\frac{r^2}{l^2}+\mathcal{O}(a^{2}Q^{2}).
\end{equation}
$Q$ is the electric charge, $a$ is a rotation parameter and is
related to the angular momentum of the solution and $l$ is related
to the cosmological constant via $\Lambda=-\frac{1}{l^{2}}$. $\beta$
is integration constants depends on the black hole charge and mass,
\begin{equation}\label{s4}
\beta=\frac{1}{3}(\frac{Q^2}{4}-M),
\end{equation}
and scalar charge $B$ related to the scalar field as,
\begin{equation}\label{s5}
\phi(r)=\pm\sqrt{\frac{8B}{r+B}}.
\end{equation}
Also one can obtain,
\begin{equation}\label{s6}
\omega(r)=-\frac{(3r+2B)a}{r^{3}},
\end{equation}
and,
\begin{equation}\label{s7}
V(\phi)=\frac{2}{l^{2}}+\frac{1}{512}\left[\frac{1}{l^{2}}+\frac{\beta}{B^{2}}+\frac{Q^{2}}{9B^{2}}\left(1-\frac{3}{2}\ln(\frac{8B}{\phi^{2}})\right)\right]\phi^{6}
+\mathcal{O}(Q^{2}a^{2}\phi^{8}).
\end{equation}
We can see that the first order of rotational parameter $a$ has no
effect on potential. So, this potential studied completely in the Ref. [1].\\
In the next section we write field equations and discuss geometric
properties of solution briefly.
\section{Field equations}
One can obtain the following independent Christophel symbols,
\begin{eqnarray}\label{s8}
\Gamma_{tt}^{r}&=&-\Gamma_{tr}^{t}=\frac{36r^{3}-9l^{2}Q^{2}r+6Bl^{2}Q^{2}\ln(r)-4Bl^{2}Q^{2}-36\beta Bl^{2}}{36l^{2}r^{2}},\nonumber\\
\Gamma_{tr}^{\psi}&=&\Gamma_{r\psi}^{t}=-\Gamma_{t\psi}^{r}=\frac{aB}{r^{2}},\nonumber\\
\Gamma_{r\psi}^{\psi}&=&-\Gamma_{\psi\psi}^{r}=r,\nonumber\\
\Gamma_{rr}^{r}&=&\frac{\mathcal{A}}{\mathcal{B}^{2}},
\end{eqnarray}
where,
\begin{eqnarray}\label{s9}
\frac{\mathcal{A}}{36l^{2}r^{3}}&=&-36r^{6}+9l^{2}Q^{2}r^{4}+4Bl^{2}(9\beta+Q^{2})r^{3}+324l^{2}a^{2}r^{2}+648Bl^{2}a^{2}r \nonumber\\
&+&288l^{2}a^{2}B^{2}-6Bl^{2}Q^{2}\ln(r)r^{3},\nonumber\\
\mathcal{B}&=&36r^{6}+9l^{2}(12\beta-Q^{2})r^{4}+4Bl^{2}(18\beta-Q^{2})r^{3}+324l^{2}a^{2}r^{2}+432Bl^{2}a^{2}r\nonumber\\
&+&144l^{2}a^{2}B^{2}-12Bl^{2}Q^{2}\ln(r)r^{3}-18l^{2}Q^{2}\ln(r)r^{4}.
\end{eqnarray}
These yield to the following Ricci scalar,
\begin{equation}\label{s10}
R=-\frac{36r^{6}-3l^{2}Q^{2}r^{4}+2Bl^{2}Q^{2}r^{3}+216Bl^{2}a^{2}r+180l^{2}a^{2}B^{2}}{6l^{2}r^{6}},
\end{equation}
which is singular at $r = 0$, and yields to $R=-\frac{6}{l^{2}}$ for
$Q=0$ and $a=0$. The Riemann and Ricci tensors also are singular at
$r = 0$ for $Q\neq0$ and $a\neq0$.\\
One of the non-vanishing components of Cotton tensor obtained as the
following,
\begin{equation}\label{s11}
C_{\psi
r\psi}=(\frac{3B}{4r^{2}}-\frac{B\ln(r)}{2r^{2}}+\frac{1}{4r})Q^{2}+(\frac{54}{r^{3}}+\frac{180B}{r^{4}}+\frac{120B^{2}}{r^{5}})a^{2}+\frac{3\beta
B}{r^{2}},
\end{equation}
which tells that, at the non conformal limit, the metric is flat.
Therefore, the rotating charged hairy black holes are geometrically
quite different from the case of the static uncharged BTZ black
hole.
\section{Horizon structure}
In the Refs. [1] and [2] horizon structure of hairy (2+1)
dimensional black holes in several special cases such as uncharged,
extremal rotating and non-extremal rotating black holes, and also
under some constraint on the black hole parameters has been studied
in details. Now we would like to discuss a more general cases
($Q\neq0$, $a\neq0$, $B\neq0$) without any conditions on the black
hole parameters. Just we take $r=z+1$ and assume $Z$ as
infinitesimal parameter ($z\ll1$ or $r-1\ll1$), therefore
$\ln(r)\approx r-1$. Under this assumption, the relation (3) gives
the following equation,
\begin{eqnarray}\label{s12}
x=r^{4}f(r)&=&r^{6}-\frac{Q^{2}}{2}r^{5}+(\frac{Q^{2}}{2}-\frac{BQ^{2}}{3}-M)r^{4}\nonumber\\
&+&\frac{B}{3}(\frac{7}{6}Q^{2}-2M)r^{3}+9a^{2}r^{2}+12Ba^{2}r+4B^{2}a^{2}=0.
\end{eqnarray}
The equation (12) has generally six solutions. Two of them are
imaginary and are not physically interesting cases. Also two of them
are negative and interpreted as naked singularities. Two remaining
solutions which are real positive are indeed inner and outer
horizons. In order to find event horizon we decompose equation (12)
as the following,
\begin{equation}\label{s13}
x=(r^{2}+c_{1})(r^{2}-c_{2})(r+c_{3})(r-c_{4}),
\end{equation}
where $r=\pm i\sqrt{c_{1}}$ are imaginary, $r=-\sqrt{c_{2}}$ and
$r=-c_{3}$ are negative solutions. Only physical solutions are
$r_{+}=\sqrt{c_{2}}$ and $r_{-}=c_{4}$ which will be interpreted as
outer and inner horizons respectively and obtained as the following,
\begin{equation}\label{s14}
c_{2}=\frac{B}{3Q^{2}}(\frac{7}{6}Q^{2}-2M)\left(-1+\sqrt{1+\frac{216a^{2}Q^{2}}{B(\frac{7}{6}Q^{2}-2M)^{2}}}\right),
\end{equation}
and,
\begin{equation}\label{s15}
c_{4}=\frac{Q^{2}}{4}\left(-1+\sqrt{1+\frac{8B}{3Q^{2}}}\right).
\end{equation}
It is clear that the $7Q^{2}\geq12M$ is crucial condition. Typical
behavior of real solutions illustrated in the Fig. 1 where $r=2$
denotes outer horizon. If we choose $Q^{2}=\frac{12}{7}M$, then
$r=0$ is event horizon.

\begin{figure}[th]
\begin{center}
\includegraphics[scale=.3]{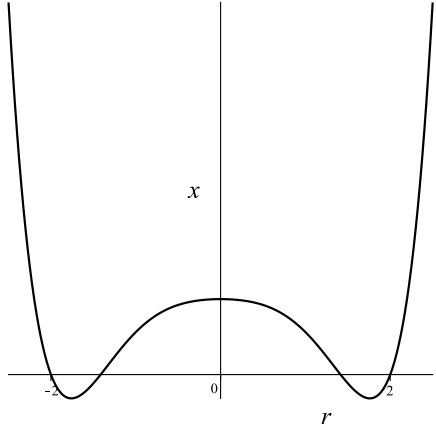}
\caption {Horizon structure}
\end{center}
\end{figure}

\section{Particle acceleration}
In order to obtain the CM energy, we should calculate the velocities
of the particles, which will be obtained as the following,
\begin{eqnarray}\label{s16}
\dot{t}&=&\frac{\omega(r)L-E}{f(r)},\nonumber\\
\dot{r}&=&\sqrt{f(r)\left(1+\frac{E^{2}}{f(r)}+\frac{\omega(r)^{2}r^{2}-f(r)}{f(r)r^{2}}L^{2}-\frac{2\omega(r)EL}{f(r)}\right)},\nonumber\\
\dot{\phi}&=&\frac{\omega(r)\left(\omega(r)L-E\right)r^{2}-f(r)L}{f(r)r^{2}}.
\end{eqnarray}
where $E$ denotes the test particle energy per unit mass and $L$
denotes the angular momentum per unit mass. We use velocity
components (16) to obtain CM energy of the two-particle collision in
the background rotating charged hairy (2+1) dimensional black hole.
It is assumed that two particles have the angular momentum per unit
mass $L_1$ and $L_2$, energy per unit mass $E_1$ and $E_2$. Moreover
we take $m_0$ as the rest mass of both particles. Then, by using the
following relation,
\begin{equation}\label{s17}
E_{CM}=\sqrt{2}m_{0}\sqrt{1+g_{\mu\nu}v_{1}^{\mu}v_{2}^{\nu}},
\end{equation}
where $v_{i}=(\dot{t}_{i}, \dot{r}_{i}, \dot{\phi}_{i})$, we can
find the CM energy of two-particle collision as the following
expression,
\begin{equation}\label{s18}
\bar{E}_{CM}=\frac{1}{f(r)r^{2}}\left(f(r)r^{2}+E_{1}E_{2}r^{2}-L_{1}L_{2}(f(r)-\omega(r)^{2}r^{2})-\omega(r)r^{2}(E_{1}L_{2}+E_{2}L_{1})-H_{1}H_{2}\right),
\end{equation}
where $\bar{E}_{CM}\equiv\frac{E_{CM}^{2}}{2m_{0}^{2}}$, and,
\begin{equation}\label{s19}
H_{i}=\sqrt{f(r)r^{2}+E_{i}^{2}r^{2}-(f(r)-\omega(r)^{2}r^{2})L_{i}^{2}-2\omega(r)r^{2}E_{i}L_{i}},
\end{equation}
with $i=1,2$. Using relations (3) and (6) in the equation (18) we
draw typical behavior of CM energy in the Fig. 2 which shows that
near the black hole horizon the CM energy takes large value which is
expected for the rotational black hole. Also dotted line of the Fig.
2, which is corresponding to static black hole ($a=0$) shows that CM
energy has finite value which is also expected.

\begin{figure}[th]
\begin{center}
\includegraphics[scale=.3]{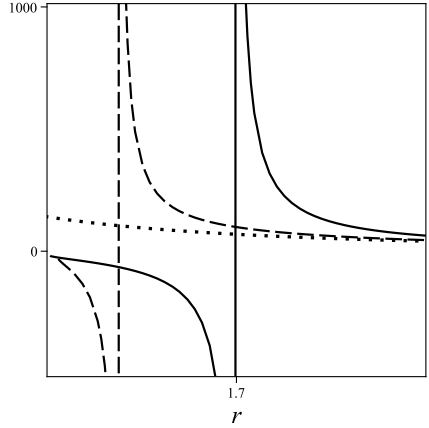}
\caption{Plots of $\bar{E}_{CM}$ in terms of $r$ by choosing
$E_{1}=E_{2}=10$, $L_{1}=-L_{2}=-10$, and $l=1$. We fix black hole
parameters as $M=5$, $Q=5$, $B=5$ and $a=1$ (solid line), $M=1$,
$Q=10$, $B=3$ and $a=1$ (dashed line), and $M=1$, $Q=10$, $B=3$ and
$a=0$ (dotted line).}
\end{center}
\end{figure}

Also we can obtain $\bar{E}_{CM}$ near the horizon analytically. In
that case we should expand $\bar{E}_{CM}$ for $r\rightarrow r_{+}$
to obtain,
\begin{equation}\label{s20}
\bar{E}_{CM}=\frac{r_{+}^{2}\left(\frac{(3r_{+}+2B)(L_{1}+L_{2})a}{r_{+}^{3}}+E_{1}+E_{2}\right)^{2}-(L_{1}E_{2}-E_{1}L_{2})^{2}}
{2r_{+}^{2}\left(\frac{(3r_{+}+2B)L_{1}a}{r_{+}^{3}}-E_{1}\right)\left(\frac{(3r_{+}+2B)L_{2}a}{r_{+}^{3}}-E_{2}\right)}.
\end{equation}
It yields to a critical angular momentum where CM energy will be
infinite,
\begin{equation}\label{s21}
L_{ci}=\frac{E_{i}r_{+}^{3}}{(3r_{+}+2B)a}.
\end{equation}
It means that the particles with the critical angular momentum
$L_{ci}$ can collide with arbitrary high CM energy near the horizon.
in the special case of vanishing scalar charge ($B=0$) we can see
that $L_{ci}\propto r_{+}^{2}$, which agrees with the results of
[26, 27]. In the Fig. 3 we draw critical angular momentum in terms
of rotational parameter $a$.

\begin{figure}[th]
\begin{center}
\includegraphics[scale=.3]{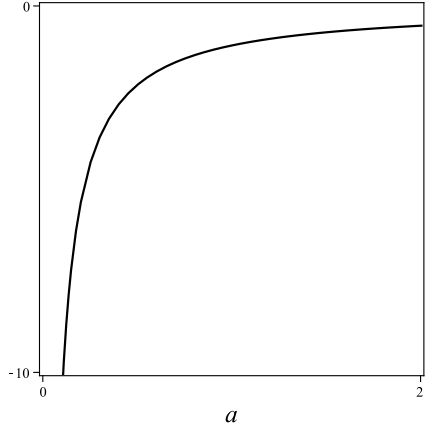}
\caption{Plots of critical angular momentum $L_{ci}$ in terms of $a$
by choosing $E_{i}=10$, $M=5$, $Q=5$, $B=5$ and $l=1$.}
\end{center}
\end{figure}

\section{Effective potential}
The effective potential is given by the following relation,
\begin{equation}\label{s22}
V_{eff}=\frac{E^{2}-R}{2},
\end{equation}
where $R=\dot{r}^{2}$ obtained by using the relation (16). In the
Fig. 4 we draw effective potential (22) for various values of the
black hole parameters. Fig. 4 (a) and (b) show that there is a
critical radius $r_{c}$ so for the $r>r_{c}$ electric charge $Q$ and
scalar charge $B$ decreased the effective potential but for the
$r<r_{c}$ the effective potential takes infinite value. Indeed, this
region is near the black hole horizon and infinite value of the
energy is expectable. In the special case of $r=r_{c}$ variation of
$Q$ and $B$ have not any effects on the effective potential. In the
Fig. 4 (c) we learn that the rotational parameter $a$ reduced the
effective potential.\\
The parameter $R$ is also useful for another reason. The particle
with the critical angular momentum may have an orbit outside the
outer horizon if,
\begin{equation}\label{s23}
O=\frac{dR}{dr}\mid_{r=r_{+}}>0.
\end{equation}
In the Fig. 5 we draw $O$ in terms of the black hole parameters.
Fig. 5 (a) tells that for the case of $a=1$ and $B=2$ the electric
charge restricted as $Q\geq5$. Fig. 5 (b) tells that for the case of
$a=1$ and $Q=5$ the scalar charge restricted as $1.5<B<26$. Fig. 5
(c) tells that for the case of $Q=5$ and $B=2$ the rotational
parameter restricted as $a\leq1$. In summary we conclude that the
condition of having an orbit outside the outer horizon is choosing
small $a$ and large $Q$ with $B_{min}<B<B_{max}$, which values of
$B_{min}$ and $B_{max}$ are depend on values of $Q$ and $a$. In
another word the electric charge increases $O$ but rotational
parameter decreases $O$, while scalar charge may increases or
decreases $O$.

\begin{figure}[th]
\begin{center}
\includegraphics[scale=.25]{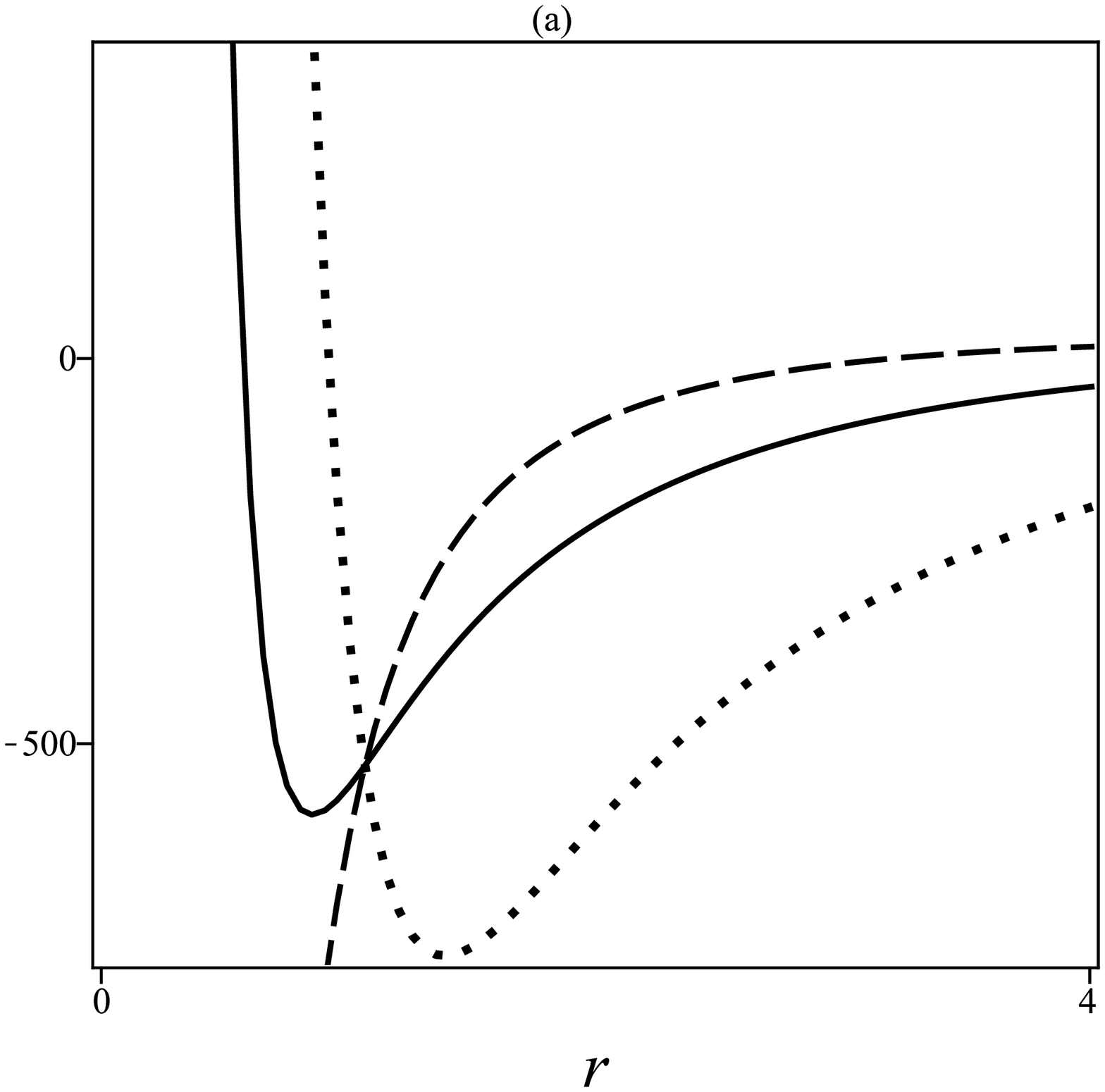}\includegraphics[scale=.25]{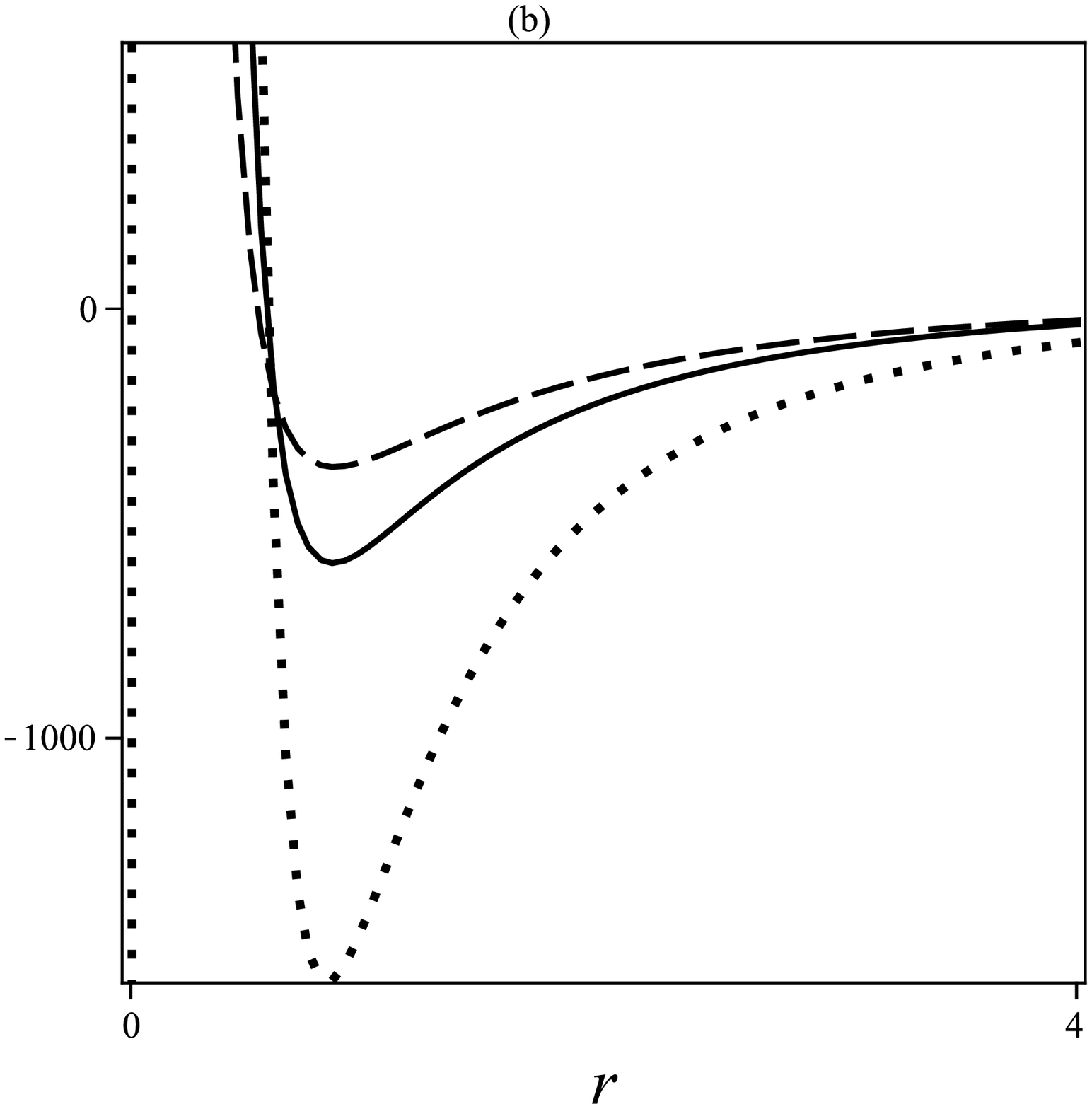}\includegraphics[scale=.25]{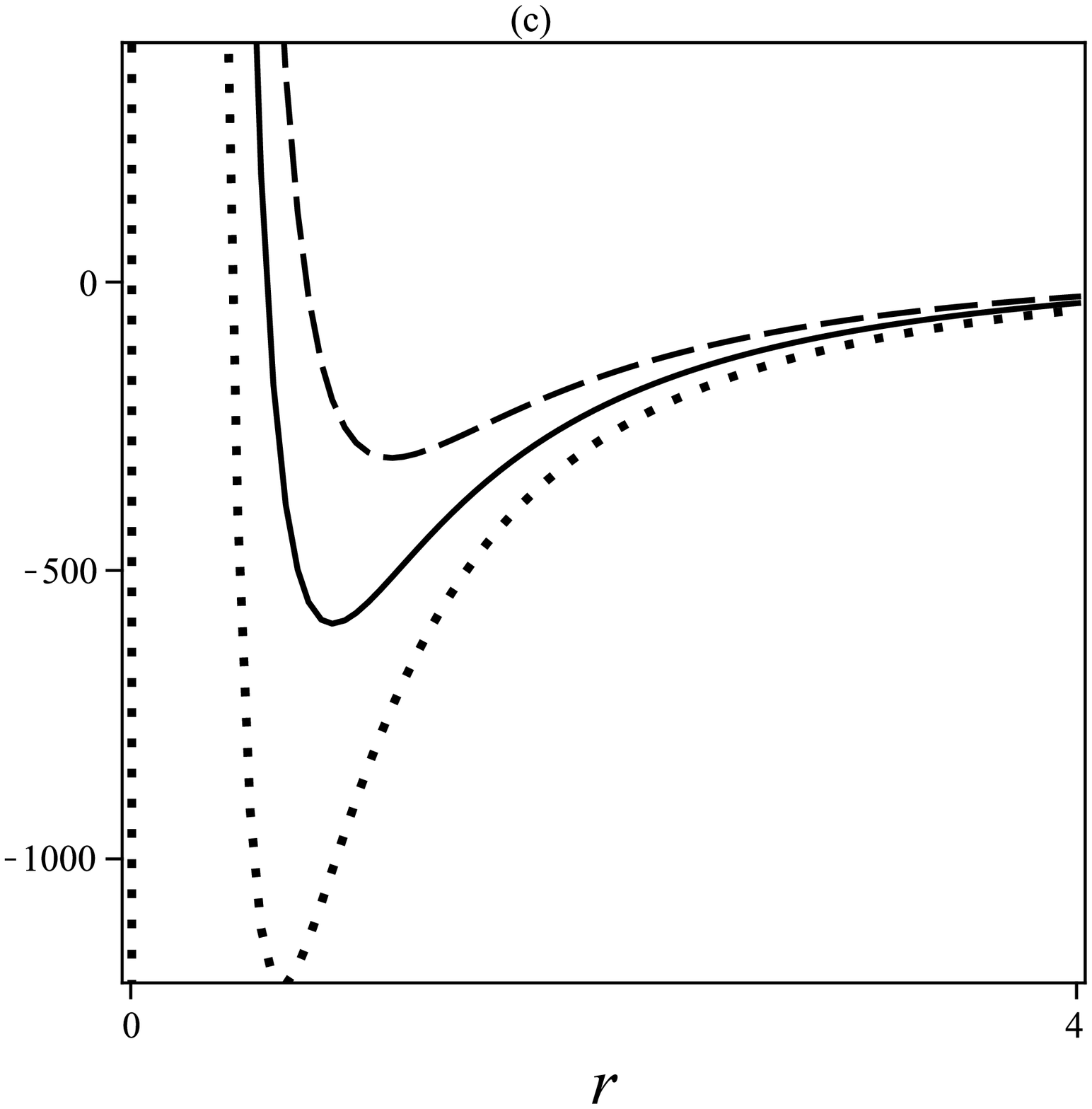}
\caption{Plot of Effective potential in terms of $r$ for $E=5$,
$L=10$, $M=5$ and $l=1$. (a) $B=1$, $a=1$, $Q=0$ (dashed line),
$Q=5$ (solid line) and $Q=10$ (dashed line). (b) $Q=5$, $a=1$, $B=0$
(dashed line), $B=1$ (solid line) and $B=5$ (dashed line). (c)
$Q=5$, $B=1$, $a=0$ (dashed line), $a=1$ (solid line) and $a=2$
(dashed line).}
\end{center}
\end{figure}

\begin{figure}[th]
\begin{center}
\includegraphics[scale=.25]{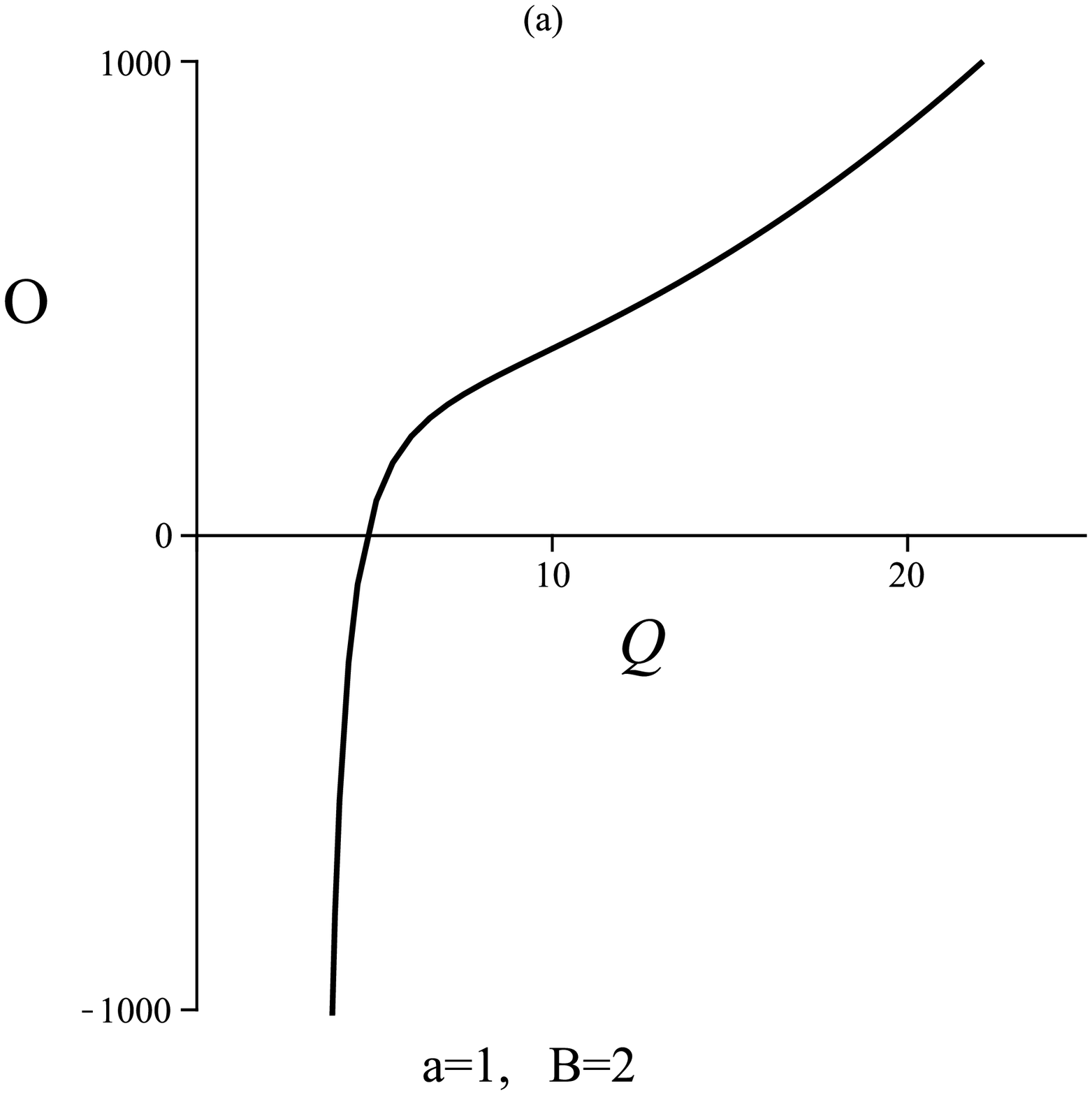}\includegraphics[scale=.25]{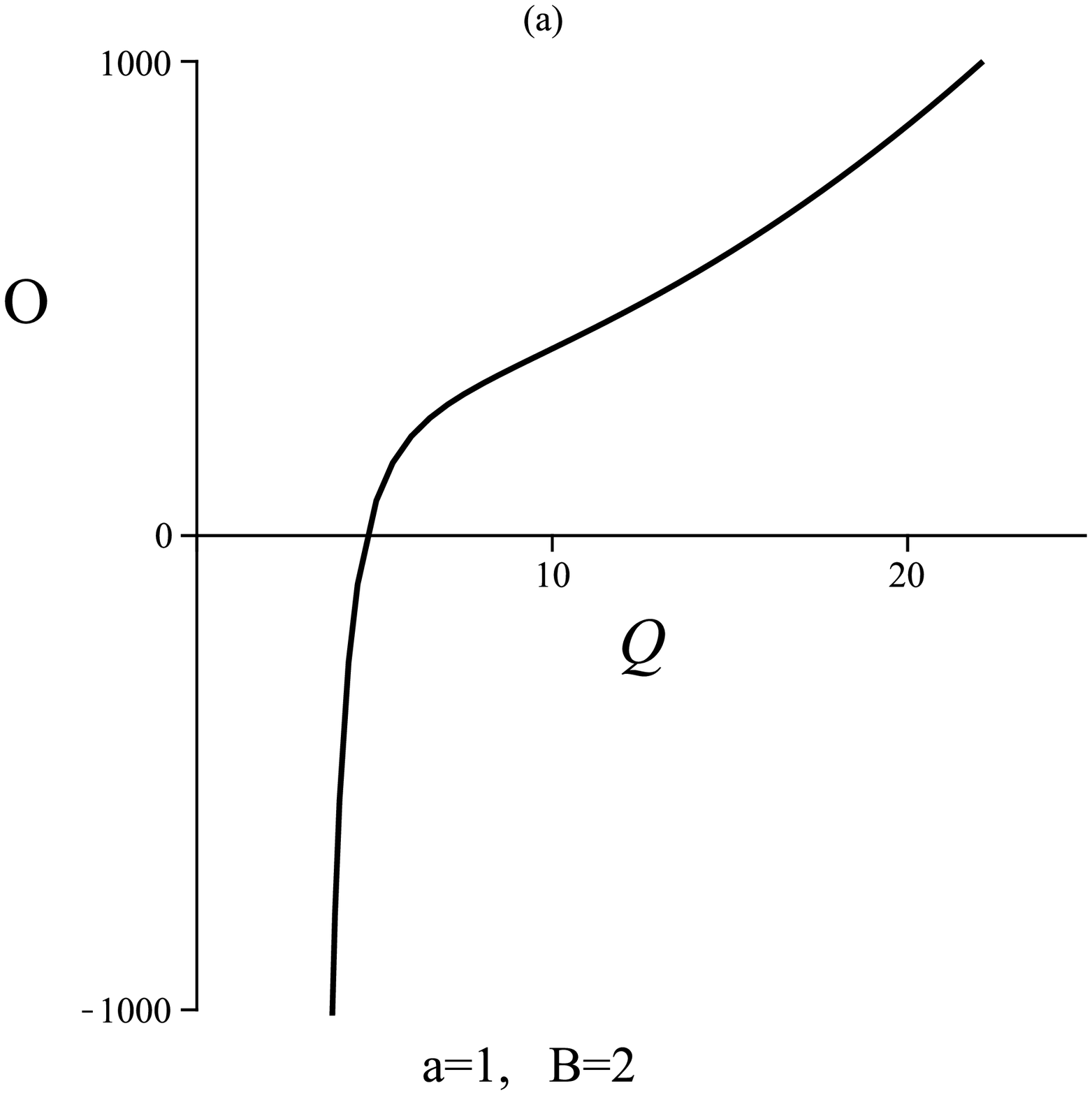}\includegraphics[scale=.25]{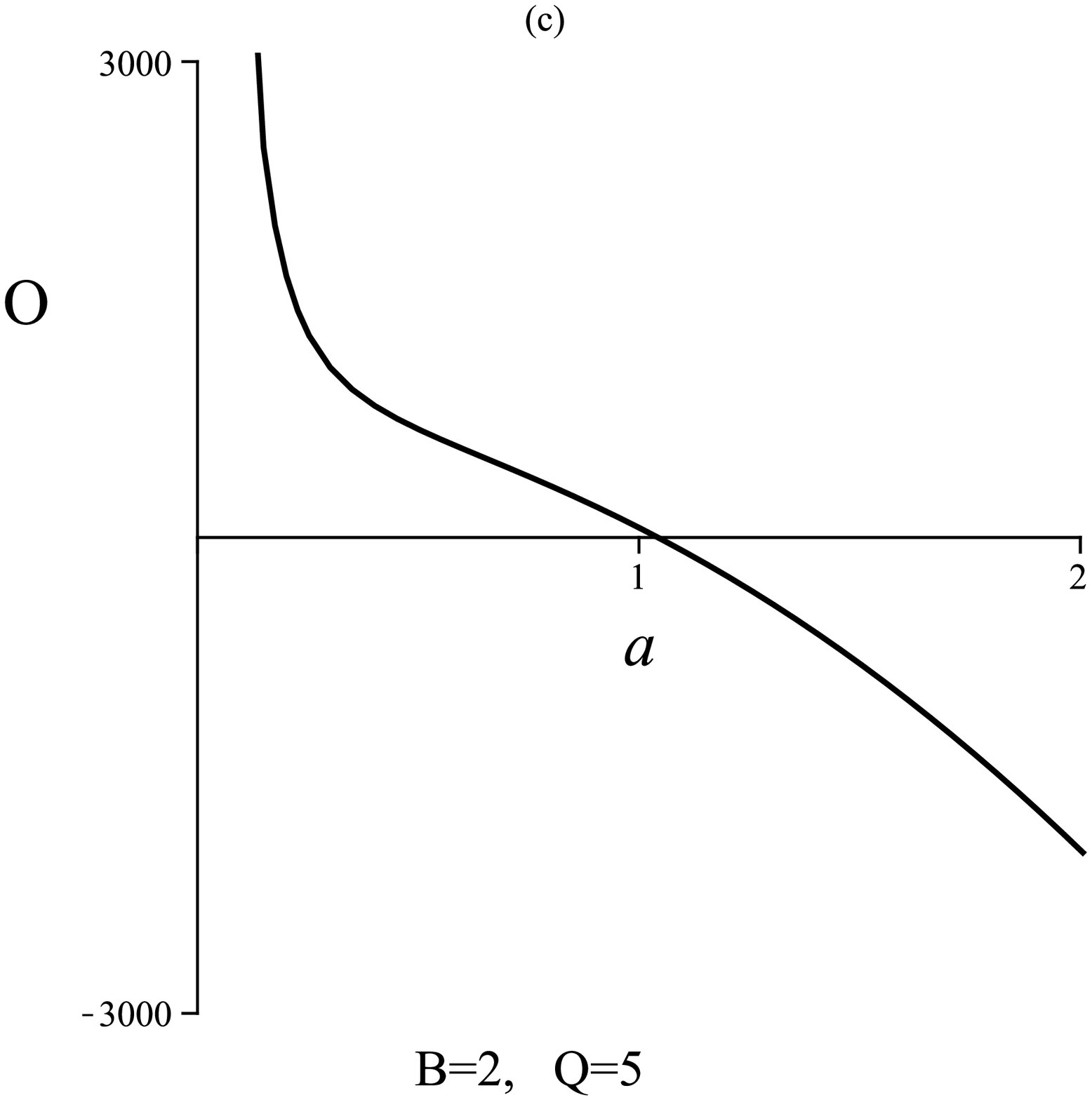}
\caption{Plot of $O$ in terms of (a) $Q$, (b) $B$ and (c) $a$ for
$E=10$, $M=5$ and $l=1$.}
\end{center}
\end{figure}
Particles on a circle orbit have the following angular momentum,
\begin{eqnarray}\label{s24}
L_{co1}&=&\frac{\omega(r)Er^{2}+r\sqrt{f(r)}\sqrt{E^{2}+\omega(r)^{2}r^{2}-f(r)}}{\omega(r)^{2}r^{2}-f(r)},\nonumber\\
L_{co2}&=&\frac{\omega(r)Er^{2}-r\sqrt{f(r)}\sqrt{E^{2}+\omega(r)^{2}r^{2}-f(r)}}{\omega(r)^{2}r^{2}-f(r)}.
\end{eqnarray}
It is easy to check that at $r=r_{+}$ where $f(r_{+})=0$ we have
$L_{ci}=L_{coi}$. This is also happen if $E^{2}\geq
f(r)-\omega(r)^{2}r^{2}$. In these cases there is no circle orbit.
Existing the circle orbit needs the angular momentum be in the
interval $L\in[L_{co2},L_{co1}]$. If we set $L_{co1}=L_{1}$, and
$L_{co2}=L_{2}-\delta$, where $0\leq\delta\leq L_{co1}-L_{co2}$,
then the CM energy in the circle orbit obtained as the following,
\begin{eqnarray}\label{s25}
\bar{E}_{CMo}&=&1+\frac{E_{1}E_{2}+\sqrt{(E_{1}^{2}+\omega(r)^{2}r^{2}-f(r))(E_{2}^{2}+\omega(r)^{2}r^{2}-f(r))}}{(\omega(r)^{2}r^{2}-f(r))}\nonumber\\
&-&\frac{\sqrt{E_{1}^{2}+\omega(r)^{2}r^{2}-f(r)}}{\sqrt{f(r)}r}\delta
+\mathcal{O}(\delta^{2}).
\end{eqnarray}
It means that the first particle is a target and the second one on
the circle orbit collide with the target. $\delta$ is the small
parameter and interpreted as the drift of the second particle from
the circle orbit.
\section{Conclusions}
In this work we constructed rotating charged black hole in (2+1)
dimensions with an scalar hair and extended recent works of Xu et
al. [1, 2]. We obtained event horizon for infinitesimal black hole
parameters and found that electric charge and scalar charge
increases size of event horizon which is agree with the results of
the Ref. [1].\\
The main part of our paper is consideration of rotating charged
hairy black hole in (2+1) dimensions as particle accelerator. We
confirmed that having arbitrary high CM energy of two colliding test
particle near the rotating black holes is universal property. Also
we found that CM energy will be finite for static black holes as
expected. We found that the angular momentum corresponding to
collision near the black hole is negative and decreased by
rotational parameter $a$. We also discussed about effective
potential and circle orbit near the black hole and found effect of
black hole parameters.\\
There are also several open problems related to these new solutions.
In the Refs. [28, 29] thermodynamics of charged and rotating hairy
black holes studied separately. Now, it is interesting to have
similar studies to the Refs. [30-32] for present background and
investigate statistical and thermodynamical quantities.\\
It is interesting to obtain quasi-normal modes [33, 34] of rotating
charged hairy black hole in (2+1) dimensions.\\
As we mentioned in the introduction there are some papers which
studied AdS/CFT correspondence in 3D hairy black hole. If this is
the case and there is a dual CFT for this background, then an
important problem will be calculation of some parameters such as
drag force and jet-quenching [35-40].

\end{document}